\documentclass[a4paper,12pt]{article}
\usepackage{amsmath}
\usepackage{amssymb}
\usepackage{latexsym}
\usepackage{enumerate}
\usepackage{bbm}
\usepackage{amsthm}
\usepackage{hyperref}
\newcommand{\be}{\begin{equation}}
\newcommand{\ee}{\end{equation}}
\newcommand{\ben}{\begin{equation*}}
\newcommand{\een}{\end{equation*}}
\newcommand{\bea}{\begin{eqnarray}}
\newcommand{\eea}{\end{eqnarray}}
\newcommand{\bean}{\begin{eqnarray*}}
\newcommand{\eean}{\end{eqnarray*}}
\newcommand{\ph}{\phantom}
\newcommand{\bsub}{\begin{subequations}}
\newcommand{\esub}{\end{subequations}}

\newcommand{\disfrac}[1][2]{\displaystyle\frac}

\newcommand{\ima}{\mathbbmtt{i}}
\newcommand{\non}{\nonumber}
\newtheorem*{mydef}{Definition}
\begin{document}

\title{\textbf{Conditional Symmetries and the Canonical Quantization of Constrained Minisuperspace Actions:\\ the Schwarzschild case}}
\vspace{1cm}
\author{\textbf{T. Christodoulakis}\thanks{tchris@phys.uoa.gr}\,, \textbf{N. Dimakis}\thanks{nsdimakis@gmail.com}\,,
\textbf{Petros A. Terzis}\thanks{pterzis@phys.uoa.gr}\\
{\it Nuclear and Particle Physics Section, Physics Department,}\\{\it University of Athens, GR 157--71 Athens}\\
\textbf{G. Doulis}\thanks{gdoulis@phys.uoa.gr}\\
{\it Department of Mathematics and Statistics, University of Otago}\\
{\it P.O. Box 56, Dunedin 9010, New Zealand}\\
\textbf{Th. Grammenos}\thanks{thgramme@civ.uth.gr}\\
{\it Department of Civil Engineering, University of Thessaly,}\\
{\it GR 383--34 Volos}\\
\textbf{E. Melas}\thanks{evangelosmelas@yahoo.co.uk}\\
{\it Logistics Department, GR 32-200, Thiva}\\
{\it Technological Educational Institution of Chalkida}\\
\textbf{A. Spanou}\thanks{aspanou@central.ntua.gr}\\
{\it School of Applied Mathematics and Physical Sciences},\\
{\it National Technical University of Athens, GR 157--80, Athens}}
\date{}
\maketitle
\begin{center}
\textit{}
\end{center}
\vspace{-1cm}
\newpage
\abstract{\textit{A conditional symmetry is defined, in the phase-space of a quadratic in velocities constrained action, as a simultaneous conformal symmetry of the supermetric and the superpotential. It is proven that such a symmetry corresponds to a variational (Noether) symmetry.The use of these symmetries as quantum conditions on the wave-function entails a kind of selection rule. As an example, the minisuperspace model ensuing from a reduction of the Einstein - Hilbert action by considering static, spherically symmetric configurations and r as the independent dynamical variable, is canonically quantized. The conditional symmetries of this reduced action are used as supplementary conditions on the wave function. Their integrability conditions dictate, at a first stage, that only one of the three existing symmetries can be consistently imposed. At a second stage one is led to the unique Casimir invariant, which is the product of the remaining two, as the only possible second condition on $\Psi$. The uniqueness of the dynamical evolution implies the need to identify this quadratic integral of motion to the reparametrisation generator. This can be achieved by fixing a suitable parametrization of the r-lapse function, exploiting the freedom to arbitrarily rescale it. In this particular parametrization the measure is chosen to be the determinant of the supermetric. The solutions to the combined Wheeler - DeWitt and linear conditional symmetry equations are found and seen to depend on the product of the two ``scale factors".}}

\numberwithin{equation}{section}

\section{Introduction}
\label{sec:intro}
In this paper we reexamine the issue of the presence of conditional symmetries in minisuperspace constrained systems. The variational symmetries approach for classical Bianchi cosmologies has been, to the best of our knowledge, initiated in \cite{Marmo}. For the case of classical and quantum cosmology, either Bianchi or higher derivatives, the first works known to us are \cite{Lamb0}, \cite{Lamb}, \cite{Cotsakis0}, \cite{Cotsakis} and \cite{tchris1} (where conditional symmetries are used). Work on the subject has been recently revived, see \cite{Vakili0}, \cite{Falciano}, \cite{Capoz} and \cite{Sarkar}. The method essentially consists in applying the standard theory of variational symmetries \cite{Olver}, \cite{Olver2} to the Lagrangian of some minisuperspace model. The usual procedure in use is to gauge fix the lapse function to some convenient value; then apply the first prolongation in the velocity phase space of a vector in the configuration space to this gauge fixed Lagrangian and demand its action to be zero. A slightly different approach using the notion of the special projective and/or homothetic group is adopted in \cite{Tsamp1}, \cite{Tsamp2} and \cite{Tsamp3}.

In our work, we do not gauge fix the lapse function. Our justification is that the presence of the lapse keeps intact the full content of the corresponding reparametrisation generator, i.e. the quadratic constraint. This is rather important in various ways, not only for the symmetries, as the present work shows, but also in solving the classical equations of motion, a task that is greatly facilitated by inserting the algebraic solution for $N$ from the quadratic constraint into the other equations of motion \cite{tchris3}, \cite{tchris4}, \cite{tchris5} and \cite{tchris6}. In the present work we start from a phase space point of view and define a conditional symmetry \cite{Kuchar} as a simultaneous conformal field of both the supermetric and the potential. We prove that this definition results in uncovering all variational symmetries. In the Appendix we give an explicit example of how the gauge fixing of the lapse function, in conjunction with the use of the standard variational condition can lead to a loss of certain existing symmetries.

We give an application of the method to the case of static, spherically symmetric configurations. A minisuperspace Lagrangian for the above family of metrics has been given in \cite{Cavaglia1}, \cite{Cavaglia2} by considering a $3+1$ decomposition along the radial coordinate $r$ which is taken to be the dynamical variable. In \cite{Vakili1}, \cite{Vakili2} Vakili has used the  symmetries of the reduced Schwarzschild action in the quantization procedure. In the present work we also adopt the quantum analogues of the linear integrals of motion as supplementary conditions imposed on the wave function. A careful examination of their role and integrability conditions leads us to the unique Casimir invariant of their algebra. This invariant is identified to the corresponding $r$ - reparametrisation generator through a particular redefinition of the lapse function.

The paper is organized as follows: In section $2$ we start with some general considerations on minisuperspace constrained actions possessing conditional symmetries (see \cite{Kuchar} for the case of full pure gravity). Subsequently, in section $3$, we give the reduced valid Lagrangian reproducing Einstein's field equations for static,  spherically symmetric configurations, and passing to its Hamiltonian formulation \cite{Sund}, we reveal the three conditional symmetries. In section $4$ we quantize the dynamical system according to Dirac's canonical quantization procedure for constrained systems \cite{Dirac}. Finally, some concluding remarks are included in the discussion.

\section{General considerations}
\subsection{Classical Treatment}
First, let us give some general considerations on conditional symmetries concerning reparametrisation invariant actions of finite degrees of freedom.
Since we would like to address collectively spatially homogeneous and/or point like geometries, we adopt as a suitable starting point the following form of the line element:
\begin{equation} \label{mastergeom}
d\,s^2=\pm N(x)^2\,d\,x^2 + g_{A B}\left(q^\alpha(x)\right)\, \omega^A_i(x^i) \omega^B_j(x^i) \, dx^i \, dx^j \quad \quad i,\,j=1,2,3.
\end{equation}
The dynamical variable $x$ may either play the role of time in the former or of the radial coordinate in the latter case, hence the $\pm$ sign in front of $dx^2$. Thus, $N(x)$, $q^\alpha(x)$ are the lapse function and the ``scale factor" components, which ought to be considered as the dependent dynamical variables of the space-time metric. Upon integration of the $x^i$ independent variables in the appropriate full action, the minisuperspace action $\mathcal{A} =\int L \, dx$ is obtained, with
\begin{equation} \label{Lag}
L= \frac{1}{2N}\, G_{\alpha\beta}(q)\, \dot{q}^\alpha \, \dot{q}^\beta - N\, V(q)
\end{equation}
\begin{equation*}
q^\alpha(x) \quad , \quad \alpha=1,...,n
\end{equation*}
It is straightforward to verify that the above action retains its form under a reparametrisation $x=f(\tilde{x})$, if one adopts the following transformations of the dependent variables $q^\alpha(x)$, $N(x)$:
\be \label{tran}
N(x) \rightarrow \tilde{N}(\tilde{x}) := N(f(\tilde{x}))\,f'(\tilde{x}) \quad , \quad q^\alpha(x) \rightarrow \tilde{q}^\alpha(\tilde{x}) := q^\alpha(f(\tilde{x})).
\ee
These transformations can be inferred from the underlying geometry \eqref{mastergeom}, as of a kinematical type, dictated by the fact that $N(x)$ appears in the line element multiplied by $dx$ while $q^\alpha(x)$ multiplied by the differentials of the coordinates which are being integrated out of the full action in order to arrive at $\mathcal{A}$. If one were presented with only the reduced Lagrangian, then \eqref{tran} should be guessed as a kind of ``gauge" transformation.  The above changes of $N(x)$ and $q^\alpha(x)$ imply that one of these variables can be prescribed at will by suitably choosing the freedom encoded in the arbitrariness of $f$; when one attempts to solve the Euler - Lagrange equations, it is both useful and instructive to let the value of $N(x)$ be defined by algebraically solving the corresponding constraint equation. If this value is substituted into the other equations concerning the $q^\alpha(x)$'s, the rank of the system is diminished by one, i.e. one can solve only for $n-1$ accelerations. This situation is reflected in the presence of two first class constraints in the Hamiltonian formulation.
\par
The momenta corresponding to the configuration variables $q^\alpha(x)$, $N(x)$ are
\begin{equation}
\pi_\alpha := \frac{\partial L}{\partial \dot{q}^\alpha}= \frac{1}{N} \, G_{\alpha\beta} \, \dot{q}^\beta \quad\quad\quad\quad \pi_N := \frac{\partial L}{\partial \dot{N}} = 0.
\end{equation}
According to Dirac's theory
\be \label{genpr}
\pi_N \approx 0
\ee
constitutes the primary constraint. The weak equality symbol $\approx$ denotes that the corresponding relation can be used only after all the Poisson brackets have been calculated. The canonical Hamiltonian is given by
\be \label{genham}
H := \pi_\gamma \, \dot{q}^\gamma - L = N \, \mathcal{H}_c=  N \, \left(\frac{1}{2} \, G^{\alpha\beta}(q) \, \pi_\alpha \, \pi_\beta + V(q)\right), \quad G_{\alpha\mu}\,G^{\mu\beta}=\delta_\alpha^\beta.
\ee
The consistency requirement that \eqref{genpr} must be preserved by the $x$ - evolution leads to the secondary constraint
\be \label{gensec}
\dot{\pi}_N:=\{\pi_N,H\}\approx 0 \Rightarrow \mathcal{H}_c \approx 0
\ee
and the algorithm is terminated since $\{\mathcal{H}_c,H\}=0$ which is stronger than what is actually needed, i.e. $\approx 0$. The two constraints are first class on account of the Poisson bracket $\{\pi_N,\mathcal{H}_c\}=0$. They represent the $x$ - reparametrisation invariance of the action and reveal $N(x)$ as a Lagrange multiplier (as pointed out above) not defined by the dynamical evolution equations involving the accelerations of the $n-1$ $q^\alpha(x)$'s.
\par
On the part of the configuration space spanned by $q^\alpha(x)$, one can define a conditional symmetry generated by a vector field $\xi$ which is a simultaneous conformal field of the metric $G^{\alpha\beta}(q)$ and the potential $V(q)$. This leads to the following:
\begin{mydef}
A vector field $\xi^\alpha$ on the configuration space, generates a conditional symmetry if
\begin{equation} \label{defcon}
\mathcal L_\xi G^{\alpha\beta} = \phi(q) \, G^{\alpha\beta} \quad, \quad \mathcal L_\xi V(q) = \phi(q)\, V(q)
\end{equation}
\end{mydef}
where $\mathcal L_\xi$ stands for the Lie derivative operator acting on the corresponding geometrical objects:
\be\label{Lie GV}
\mathcal L_\xi G^{\alpha\beta} := G^{\alpha\beta}_{,\rho} \, \xi^\rho - G^{\rho\beta}\, \xi^\alpha_{,\rho} - G^{\alpha\rho}\, \xi^\beta_{,\rho} \quad, \quad \mathcal L_\xi V(q) := V(q)_{,\rho}\, \xi^\rho.
\ee

To each of the existing conditional symmetries one can correspond a phase space quantity
\be \label{charge}
Q_I := \xi_I^{\alpha} \pi_\alpha.
\ee
As one can easily check, the Poisson brackets of \eqref{charge} with the Hamiltonian vanish on the constrained surface on account of \eqref{gensec}:
\begin{equation} \label{consym}
\{Q_I, N \mathcal{H}_c\} =- N \left(\frac{1}{2} \left(\mathcal{L}_{\xi_I} G^{\alpha\beta}\right) \pi_\alpha \pi_\beta + \mathcal{L}_{\xi_I} V \right) =-N \, \phi_I (q) \, \mathcal{H}_c \approx 0.
\end{equation}

Therefore, by virtue of \eqref{consym} the quantities \eqref{charge} are constants of motion
\be \label{constant}
Q_I = \kappa_I.
\ee
As it is well known, the totality of these charges forms a Lie algebra characterized by some structure constants, say $C^M_{IJ}$
\be \label{classalg}
\{Q_I,Q_J\}=C^M_{IJ} \, Q_M.
\ee

The equation $\{Q_I, H\}\approx 0$ is the definition for conditional symmetries used in \cite{Kuchar}. Our definition of conditional symmetries \eqref{defcon} is of course equivalent to the aforesaid definition, so it is a matter of taste which one should be preferred over the other. Moreover, \eqref{defcon} are somewhat different than the conditions used in the earlier works cited in the introduction, in the sense that we allow the $\xi_I$'s to be conformal fields of the supermetric and not Killing fields. This is justified by the weak vanishing of the superhamiltonian $H$ resulting from \eqref{gensec}.

A further important property emanating from this definition is the fact that under a scaling of $G^{\alpha\beta}$ and $V(x)$ by, say $\omega(q)$, the conserved charges \eqref{charge} retain their nature and, more usefully, remain form invariant. To see this, let us act with the same $\xi_{(I)}$ on $\bar{G}^{\alpha\beta}=\omega(q)\,G^{\alpha\beta}$ and $\bar{V}=\omega(q)\,V(q)$:
\be
\mathcal L_\xi \left(\omega(q)V\right) =\omega(q)\, \mathcal L_\xi V + \mathcal L_\xi \omega(q)\,V = \left( \phi + \frac{\xi^\alpha\, \omega_{,\alpha}}{\omega} \right) \omega(q)V =: \Phi(q)\left(\omega(q)\,V\right)
\ee
and likewise for $\bar{G}^{\alpha\beta}$. The above equation shows that if we form the new quantities $\bar{Q}_I=\xi^\alpha_I\,\bar{\pi}_\alpha$, these remain constants of motion, since their Poisson bracket with the $\bar{\mathcal{H}}_C$ still vanishes weakly.

The scaling by $\omega$ can be achieved by rescaling $N$, i.e. $N=\omega(q) \bar{N}$, as it can be seen by the form of \eqref{genham}. A particularly interesting and useful choice of $\omega(q)$ is the one that makes $\Phi(q)$ equal to zero. This $\omega(q)$ can be evaluated with the aid of the second of \eqref{Lie GV}:
\be
\mathcal{L}_\xi V=\phi(q) V\Rightarrow \xi^\alpha\,V_{,\alpha}=\phi(q)\, V\Rightarrow \phi(q)=\frac{\xi^\alpha\,V_{,\alpha}}{V}.
\ee
Use of this $\phi$ into the definition of $\Phi(q)$ leads to a linear partial differential equation for $\omega$ which can always be solved.

An obvious solution is $\omega=\lambda V^{-1}$ (with $\lambda$ being a constant) which makes the potential $\bar{V}$ constant. Furthermore, in this case the $\xi_{I}$'s are turned into Killing fields of $\bar{G}^{\alpha\beta}$ rather than conformal fields. It is in this particular parametrization for the lapse that a connection can be established with the usual variational (Noether) symmetries approach: The Killing fields $\xi_I$, with the appropriate prolongation, provide the fields $X_I$ which are the generators of the corresponding symmetries, i.e. they satisfy $pr^{(1)}X_I (\bar{L})+\bar{L} \text{Div}\Xi_I = \text{Div} F_I$ \cite{Olver2} (with $\Xi_I$ being the coefficient of the partial derivative of the independent variable $x$ and $F_I(q^\alpha,\,x)$ being an arbitrary function). This is rather important, since if one applies the above condition to another lapse parametrization in which the potential is not constant and the $\xi_I$'s are not Killing vector fields, one is bound to loose some symmetries, if not all as the Appendix shows for the case of the specific example of section $3$. On the other hand, our definition of symmetries \eqref{defcon}, based on the phase space and inspired by the weak vanishing of the superhamiltonian, can be applied to any conformal gauge. Therefore we have the following theorem: \\ \\
\textbf{Theorem. } \textit{For the case of constrained systems described by \eqref{Lag}    , the conditional symmetries are equivalent to Noether symmetries.}
\\

If one wished, one could also work in the velocity phase space and in an arbitrary parametrization of $N(x)$. The price one would have to pay is to accordingly modify the variational symmetry condition by allowing the right hand side to be expressed as a multiple of the constraint Euler - Lagrange equation with respect to $N$. Let us start with
\begin{align}
& X :=\xi^\alpha(q) \frac{\partial}{\partial q^\alpha}, \quad pr^{(1)} X  := X + \frac{d \xi^\alpha(q)}{d x} \frac{\partial}{\partial \dot{q}^\alpha}
\end{align}
\begin{mydef}
A vector field $X=\xi^\alpha(q)\,\partial_\alpha$ generates a conditional symmetry if
\begin{align}\label{Lagcon}
pr^{(1)}X(L)=\phi(q)\,N\,EL(N)
\end{align}
where $EL(N)$ is the Euler-Lagrange equation with respect to $N$.
\end{mydef}
Using the above definition we can easily compute the action of $pr^{(1)}X$ on \eqref{Lag}
\begin{align}
pr^{(1)}X(L)=\phi(q)\,N\,EL(N)\Rightarrow\non\\
\frac{1}{2 N}\left(\mathcal{L}_\xi G_{\alpha\beta}\right)\dot{q}^\alpha \, \dot{q}^\beta-N \mathcal{L}_\xi V=-\phi(q)\,N\,\left(\frac{1}{2N^2}\, G_{\alpha\beta}\, \dot{q}^\alpha \, \dot{q}^\beta + \, V\right) \Rightarrow\non\\ \label{Lagcon2}
\frac{1}{2N}\left(\mathcal{L}_\xi G_{\alpha\beta}+\phi(q)\,G_{\alpha\beta}\right)\dot{q}^\alpha\,\dot{q}^\beta-N\,\left(\mathcal{L}_\xi V-\phi(q)\,V\right)=0 \quad \forall \, \dot{q}^\mu
\end{align}
which, in order to be an identity for all $\dot{q}^\mu$  leads to the conditions \eqref{defcon} thus proving the equivalence of the two definitions. This can also be visualized as an enlargement of the relevant vector $X$ by adding the term $-\phi(q) N\frac{\partial}{\partial N}$, making manifest the freedom to arbitrary rescale $N$.

\subsection{Quantum Treatment}
For the canonical quantization of the above system, we adopt the point of view of promoting to operators not only $\mathcal{H}_c$, but also the generators of the conditional   symmetries $Q_I$. In the Schr\"{o}dinger picture the first class constraints become the quantum conditions:
\begin{align} \nonumber
& \widehat{\pi}_N \Psi(q,N) := - \ima \frac{\partial}{\partial N} \Psi(q,N) = 0 \Rightarrow \Psi \equiv \Psi(q) \\ \label{quantcon}
& \widehat{\mathcal{H}}_c \Psi = 0 \Rightarrow \left[- \frac{1}{2\mu} \partial_\alpha (\mu \, G^{\alpha\beta} \partial_\beta) + V(q) \right] \Psi = 0
\end{align}
where $\mu$ is a suitable measure which must transform as a scalar density; under $\mu$ the quantum quadratic constraint \eqref{quantcon} is Hermitian. This form is the most general second order scalar operator, except for a possible multiple of the Ricci scalar, $\mathcal{R}$, corresponding to $G_{\alpha\beta}$.

A natural scalar density emerging from the geometric structure of the configuration space is $\sqrt{|\det{G_{\alpha\beta}}|}=\sqrt{|G|}$. It is reasonable to limit the possible measures to be inferred from the classical geometrical structure, as $\mu=f(\sqrt{|G|})$. However, this choice of $\mu$ in conjunction to the scalar density transformation law for $\mu$ and $\sqrt{|G|}$ leads to the equation $\bar{\mu}(\sqrt{|G|}J)=\mu(\sqrt{|G|})J$, which must hold for any transformation $q^\alpha = q^\alpha(\tilde{q}^\beta)$ with $J=\det \frac{\partial q}{\partial \tilde{q}}$. The only solution is therefore $\mu=e^{\lambda(q^\alpha)} \sqrt{|G|}$, with $\lambda$ any scalar. Thus, one must essentially select the Laplacian based on the configuration space metric, if $\lambda(q^\alpha)=const.$, and the addition should then be exactly $\frac{n-2}{4(n-1)} \mathcal{R}$ in order for the operator to be conformally covariant \cite{ChrisZan}.
\par
The conditional symmetries \eqref{charge}, \eqref{constant} must also be turned into Hermitian operators (see \cite{tchris2}), as :
\begin{subequations} \label{qdef}
\begin{align}
& \label{qlin}\widehat{Q}_I \Psi := - \frac{\ima}{2\mu} \left(\mu \, \xi_I^\alpha\, \partial_\alpha + \partial_\alpha \, \mu\, \xi_I^\alpha \right) \Psi \\
& \label{qeigen} \widehat{Q}_I\Psi = \kappa_I \Psi.
\end{align}
\end{subequations}
If we bear in mind that the classical quantities $Q_I$ generate symmetries, it is mandatory, in order to retain their geometrical character, to demand that they act as derivatives on the wave function $\Psi$. In other words, we must have the momentum operators acting on the far right side, which translates into the condition $\frac{\partial_\alpha(\mu\, \xi_I^\alpha)}{\mu}=0$. This equation can be satisfied in the particular conformal parametrization $N=\frac{\bar{N}}{V}$, since then
\begin{align}
\frac{\partial_\alpha(\mu\, \xi_I^\alpha)}{\mu}=\frac{\partial_\alpha\left(\exp\left(\lambda\right) |\bar{G}|^{1/2} \, \xi_I^\alpha\right)}{\exp\left(\lambda\right)|\bar{G}|^{1/2}}= \xi_I^\alpha \partial_\alpha \lambda + \xi^\alpha_{I;\alpha}=0 \Rightarrow
\xi_I^\alpha \partial_\alpha \lambda =0.
\end{align}
The term $\xi^\alpha_{I;\alpha}$ is zero if evaluated with the metric $\bar{G}_{\alpha\beta}$ in which the $\xi_I$'s are Killing fields.

The maximum number of $Q_I$'s is $\frac{n(n+1)}{2}$. If the actual existing symmetries are more than or equal to $n$, the above equation fixes $\lambda$ to a constant value. For a lesser number of symmetries, some non trivial $\lambda(q)$ may be allowed.

Upon quantization it is reasonable to assume that a quantum algebra isomorphic to the classical one holds:
\be
\left\{\cdot,\cdot\right\}\mapsto \frac{-\ima}{\hbar}\,\left[\cdot,\cdot\right]
\ee
which turns \eqref{classalg} into
\be\label{qiom}
\left[\widehat{Q}_I,\widehat{Q}_J\right]=\ima\,\hbar\,C^M_{\ph{M}IJ}\,\widehat{Q}_M\Rightarrow\left[\widehat{Q}_I,\widehat{Q}_J\right]=\ima \,C^M_{\ph{M}IJ}\,\widehat{Q}_M
\ee
where in the last equality we used units in which $\hbar=1$.
At first sight equation \eqref{qiom} in conjunction with \eqref{qdef} seems to impose certain restrictions on the form of the $\xi$'s, the constants $\kappa_I$ and an initially arbitrary measure $\mu(q^\alpha)$ (although we have already committed ourselves to $\mu\propto\sqrt{|G|}$). In order to begin exploring these restrictions, we act with \eqref{qiom} on $\Psi(q^\alpha)$ and we use \eqref{qeigen}
\begin{align}\label{1st int}
\left[\widehat{Q}_I,\widehat{Q}_J\right]\,\Psi(q^\alpha)&=\ima \,C^M_{\ph{M}IJ}\,\widehat{Q}_M\,\Psi(q^\alpha)\nonumber\Rightarrow\\
\left(\kappa_I\,\kappa_J-\kappa_J\,\kappa_I\right)\,\Psi(q^\alpha)&=\ima \,C^M_{\ph{M}IJ}\,\kappa_M\,\Psi(q^\alpha)\nonumber\Rightarrow\\
C^M_{\ph{M}IJ}\,\kappa_M&=0
\end{align}
The above equation must be interpreted in a recursive and exhaustive manner: It must be checked for all subalgebras since they will correspond to different admissible choices of subsets of the $Q_I$'s, $I \subset \{1,\ldots,\frac{n(n+1)}{2}\}$. The number of essential constants for the underlying geometry \eqref{mastergeom} can provide us with a lower bound of the dimension of the subalgebras to be selected: If \eqref{mastergeom} depends on some constants $\kappa_1$, $\kappa_2$, ..., one can always try to find a coordinate transformation of $x,\, x^i$ that absorbs some (or all) of the constants. If a constant cannot be absorbed by such a transformation we call this constant essential. From this definition it is apparent that the constants not appearing in \eqref{1st int} are eligible candidates for representing the geometry. Such constants will belong to the centre of the corresponding subalgebra.

Further restrictions, induced by \eqref{qiom}, could come to life when we expand this equation taking into account \eqref{qlin}. The left hand side of \eqref{qiom}, then reads
\begin{align}\label{lhs}
l.h.s&=\left[-\ima\,\xi^{\ph{I}\alpha}_I\,\partial_\alpha+\frac{-\ima}{2\,\mu(q^\nu)}\left(\mu(q^\nu)\,\xi^{\ph{I}\alpha}_I\right)_{,\alpha},
-\ima\,\xi^{\ph{J}\beta}_J\,\partial_\beta+\frac{-\ima}{2\,\mu(q^\nu)}\left(\mu(q^\nu)\,\xi^{\ph{J}\beta}_J\right)_{,\beta}\right]\nonumber\\
&=-\left[\xi^{\ph{I}\alpha}_I\,\partial_\alpha,\xi^{\ph{J}\beta}_J\,\partial_\beta\right]-
\left[\xi^{\ph{I}\alpha}_I\,\partial_\alpha,\frac{\left(\mu(q^\nu)\,\xi^{\ph{J}\beta}_J\right)_{,\beta}}{2\,\mu(q^\nu)}\right]
+
\left[\xi^{\ph{J}\beta}_J\,\partial_\beta,\frac{\left(\mu(q^\nu)\,\xi^{\ph{I}\alpha}_I\right)_{,\alpha}}{2\,\mu(q^\nu)}\right].
\end{align}
The first commutator of \eqref{lhs} can be evaluated quite easily if we expand \eqref{classalg}
\be\label{comxi}
\xi^{\ph{J}\beta}_J\,\xi^{\ph{I}\alpha}_{I\ph{a},\beta}-\xi^{\ph{I}\beta}_I\,\xi^{\ph{J}\alpha}_{J\ph{a},\beta}=C^M_{\ph{M}IJ}\,\xi^{\ph{M}\alpha}_M,
\ee
while the last two commutators of \eqref{lhs} combine to
\be
\frac{\mu(q^\nu)_{,\beta}}{2\,\mu(q^\nu)}\left(\xi^{\ph{I}\alpha}_I\,\xi^{\ph{J}\beta}_{J\ph{b},\alpha}-\xi^{\ph{J}\alpha}_J\,\xi^{\ph{I}\beta}_{I\ph{b},\alpha}\right)+\frac{1}{2}\left(\xi^{\ph{I}\alpha}_I\,\xi^{\ph{J}\beta}_{J\ph{b},\alpha\beta}-\xi^{\ph{J}\alpha}_J\,\xi^{\ph{I}\beta}_{I\ph{b},\alpha\beta}\right).
\ee
The right hand side of \eqref{qiom} reads
\begin{align}\label{rhs}
r.h.s.&=\ima\,C^M_{\ph{M}IJ}\left(-\ima\,\xi^{\ph{M}\alpha}_M\,\partial_\alpha+\frac{-\ima}{2\,\mu(q^\nu)}\left(\mu(q^\nu)\,\xi^{\ph{M}\alpha}_M\right)_{,\alpha}\right)\nonumber\\
&=C^M_{\ph{M}IJ}\,\xi^{\ph{M}\alpha}_M\,\partial_\alpha+
\frac{C^M_{\ph{M}IJ}}{2\,\mu(q^\nu)}\left(\mu(q^\nu)\,\xi^{\ph{M}\alpha}_M\right)_{,\alpha}.
\end{align}

Gathering the above results we get the integrability condition
\begin{align}
\mu(q^\nu)_{,\beta}\left(\xi^{\ph{I}\alpha}_I\,\xi^{\ph{J}\beta}_{J\ph{b},\alpha}-\xi^{\ph{J}\alpha}_J\,\xi^{\ph{I}\beta}_{I\ph{b},\alpha}\right)
+\mu(q^\nu)\left(\xi^{\ph{I}\alpha}_I\,\xi^{\ph{J}\beta}_{J\ph{b},\alpha\beta}-\xi^{\ph{J}\alpha}_J\,\xi^{\ph{I}\beta}_{I\ph{b},\alpha\beta}\right)\nonumber \\ +C^M_{\ph{M}IJ}\left(\mu(q^\nu)\,\xi^{\ph{M}\alpha}_M\right)_{,\alpha}=0 \label{2nd int A}
\end{align}
and using once more \eqref{comxi} we arrive at
\begin{align}\label{2nd int}
C^M_{\ph{M}IJ}\,\nabla\xi_M+\left(\mathcal{L}_I\,\nabla\xi_J-\mathcal{L}_J\,\nabla\xi_I\right)=0,
\end{align}
where $\nabla \xi_I \equiv \partial_\alpha\xi_I^\alpha$. Surprisingly enough, the above relation is nothing but an identity that can be derived by taking the divergence of \eqref{comxi}, leaving us with only \eqref{1st int}.

Finally, let us point out that if someone wished to investigate the possibility that the quantum algebra is not isomorphic to the classical algebra, thus relaxing \eqref{qiom} and just evaluating the commutator on the wave function $\Phi(q^\alpha)$, then one would obtain a combined relation of the form
\begin{equation}
\ima\,C_{\ph{M}IJ}^M \kappa_M -\frac{1}{2}\left(\mathcal{L}_{\xi_I} \nabla \xi_J - \mathcal{L}_{\xi_J} \nabla  \xi_I+C_{\ph{M}IJ}^M\,\nabla \xi_M\right)= 0
\end{equation}
which, however, is again reduced to \eqref{1st int}. This establishes \eqref{1st int} as the only constraint on the classical quantities, even without the assumption of the quantum algebra being isomorphic to the classical algebra.

These constraints among the $\kappa_L$'s prohibit the simultaneous realization of all $\widehat{Q}_L$'s as eigen-operators. Usually one adopts the maximal Abelian subgroup along with the Casimir invariant as consistent conditions on the wave function. However, choices of non-Abelian subgroups may also be consistent \cite{Dirac2}. Further investigation can be made for particular systems as in the example below.

\section{Hamiltonian formulation of static, spherically symmetric geometries}
Our starting point is the static, spherically symmetric line element
\be
 \label{metric}
  ds^2 = - a(r)^2 dt^2 + n(r)^2 dr^2 +  b(r)^2 (d\theta^2+ \sin^2\theta\,d\phi^2 ),
\ee
where we have changed the notation of the lapse function from $N(x)$ to $n(r)$.

In the usual ADM $3+1$ decomposition one foliates space-time in $t$-hypersurfaces and the coefficient of $dt^2$ is the lapse function. Here we adopt a $3+1$ foliation in the $r$ coordinate and therefore the role of the lapse is attributed to the coefficient of $dr^2$ in the line element. Thus we consider $n(r)$ in \eqref{metric} to be the $r$-lapse function while $a(r)$ and $b(r)$ are the dependent ``dynamical" variables on the $r$-hypersurface. Previous authors use also a ``shift" term $2 B(r)\,dr\,dt$. However, this is not relevant for the following discussion, as $B(r)$ does not enter the Einstein tensor while it can be absorbed by a time redefinition of the form $t=\tilde{t}+\int\! \frac{B(r)}{a(r)^2} dr$.

The Einstein - Hilbert action $A_{E-H} = \int \! \sqrt{-g}\, R\, d^4 x$ for the geometries \eqref{metric} leads to the reduced action $\mathcal{A}=\int L dr$ with the following Lagrange function $L(a,b,a',b',n)$:
\be
 \label{lagrangian}
  L = 2 \, a \, n + \frac{4 \, b \, a' \, b'}{n} + \frac{2 \, a \, b'^2}{n},
\ee
where $'$ denotes differentiation with respect to the spatial coordinate $r$. It is easy to verify that the Euler - Lagrange equations obtained from \eqref{lagrangian} are identical to Einstein's equations $\mathcal{G}_{\mu\nu}=0$ for the line element \eqref{metric}.
\par
In order to proceed with the Hamiltonian formalism we calculate the conjugate momenta,
\bea
 \label{conj_momenta}
  \begin{aligned}
   \pi_n &= \frac{\partial \mathcal{L}}{\partial n'} = 0, \\
   \pi_a &= \frac{\partial \mathcal{L}}{\partial a'} = \frac{4 b \, b'}{n}, \\
   \pi_b &= \frac{\partial \mathcal{L}}{\partial b'} = \frac{4 b \, a'}{n} + \frac{4 \, a \, b'}{n}.
  \end{aligned}
\eea
Obviously, $\pi_n$ is a primary constraint. The Legendre transformation leads to the Hamiltonian
\ben
  H = n \, \mathcal{H}_c,
\een
where
\be
 \label{hamiltonian}
  \mathcal{H}_c = - 2 \, a - \frac{a \, \pi_a^2}{8 \, b^2} + \frac{\pi_a \, \pi_b}{4 \, b}.
\ee
The preservation of the primary constraint $\pi_n$ in the $r-$evolution, i.e.
\ben
 \pi'_n = \{\pi_n, H\} \approx 0,
\een
leads to the secondary constraint
\be
 \label{quadratic_const}
  \mathcal{H}_c \approx 0.
\ee

The minisuperspace metric inferred from \eqref{hamiltonian} is
\be \label{supmet}
G^{\alpha\beta} =
\begin{pmatrix}
-\frac{a}{4 b^2} & \frac{1}{4 b} \\ \\
\frac{1}{4 b} & 0
\end{pmatrix} .
\ee
Our definition \eqref{defcon} is fulfilled by the following three conformal Killing fields of both $G_{\alpha\beta}$ and the potential $V=-2a$:
\bea
 \label{xiI}
  \begin{aligned}
  \xi_1 = (-a,b), \quad
   \xi_2 = \left(\frac{1}{a \, b},0\right), \quad
   \xi_3 = \left(- \frac{a}{2 \, b},1\right)
  \end{aligned}
\eea
which, contracted with $\left(\pi_a,\pi_b\right)$, provide us with the three integrals of motion:
\bea
 \label{int_motion}
  \begin{aligned}
   Q_1 = -a \, \pi_a + b \, \pi_b, \quad
   Q_2 = \frac{\pi_a}{a \, b}, \quad
   Q_3 = - \frac{a \, \pi_a}{2 \, b} + \pi_b.
  \end{aligned}
\eea

We calculate the Poisson brackets of these conserved quantities with the canonical Hamiltonian $H$ and the Poisson algebra that they satisfy:
\begin{subequations}
\begin{align} \label{poisson_int_motion}
    & \{Q_1, H \} = n \mathcal{H}_c, \quad
    \{Q_2 ,H \} = - \frac{n}{a^2 \, b}\,\mathcal{H}_c, \quad
    \{Q_3, H\} = \frac{n}{2 \, b}\,\mathcal{H}_c,  \\ \label{poisson_int_motion2}
    & \{Q_1 , Q_3 \} = Q_3, \quad
    \{Q_2 , Q_1 \} = Q_2, \quad
    \{Q_3 , Q_2 \} = 0.
\end{align}
\end{subequations}
As expected from the discussion of the general case \eqref{consym}, the Poisson brackets \eqref{poisson_int_motion} are weakly vanishing on the constraint surface $\mathcal{H}_c\approx 0$ and therefore the three $Q_I$'s are constants of motion.

At this point it is interesting, and useful for what follows in the quantization, to adopt a new parametrization of the lapse $n(r)=\frac{\bar{n}(r)}{2a(r)}$ which makes the potential constant as explained in the previous section. The Lagrangian and the corresponding Hamiltonian are now given by
\be
 \label{lagham}
  \bar{L} = \bar{n} + \frac{8\, a\, b\, a' \, b'}{\bar{n}} + \frac{4 \, a^2 \, b'^2}{\bar{n}} \quad\quad, \quad\quad \bar{H} =\bar{n}\bar{\mathcal{H}}_c= \bar{n}\frac{1}{2a}\mathcal{H}_c.
\ee
If the value of $\bar{n}$ specified by the constrained equation is substituted into the Euler - Lagrange equations for $a(r)$ and $b(r)$ the system can be solved for only one acceleration, say $a''(r)$, and the general solution of the entire system is:
\be \label{classol}
\bar{n}(r)= 2\, c\, b'(r)  \quad , \quad a(r) =c \sqrt{1-\frac{2M}{b(r)}} ,
\ee
where the constants of integration have been rearranged so that the ensuing line element
\be \label{linel2}
ds^2 = -c^2\left(1-\frac{2M}{b(r)}\right)dt^2 + \left(1-\frac{2M}{b(r)}\right)^{-1}b'(r)^2 dr^2 + b(r)^2 (d\theta^2+\sin^2\theta\,d\phi^2  )
\ee
bears the closest possible resemblance to the standard form of the Schwarzschild metric, while the presence of the arbitrary function $b(r)$ reflects the $r$ reparametrisation covariance of the system.

The new supermetric is now given by
\be \label{gbar}
\bar{G}^{\alpha\beta} =
\begin{pmatrix}
-\frac{1}{8 b^2} & \frac{1}{8 a\, b} \\ \\
\frac{1}{8 a\, b} & 0
\end{pmatrix} ,
\ee
and the fields \eqref{xiI} are now turned into Killing fields of $\bar{G}^{\alpha\beta}$; they are also, trivially, symmetries of the constant potential $\bar{V}=1$ as well. The algebra satisfied by the new quantities $\bar{Q}_I=\xi_I^\alpha \bar{\pi}_\alpha$ and $\bar{\mathcal{H}}_c$ can be easily seen to be
\be
\{\bar{Q}_1, \bar{\mathcal{H}}_c\} = 0, \quad
\{\bar{Q}_2 ,\bar{\mathcal{H}}_c\} = 0, \quad
\{\bar{Q}_3, \bar{\mathcal{H}}_c\} = 0.
\ee
Therefore the quantities $\bar{Q}_I$ still remain constants of motion. It is in this conformal gauge that one can explicitly verify that the prolonged $\xi_I$'s, $X_I$, are satisfying a form of the standard condition of variational symmetries $pr^{(1)}X_I(\bar{L})=0$, thus ensuring that they are generators of Noether symmetries. It is important to note that if one had inserted the $X_I$'s into the full relation $pr^{(1)}X (L)+L \text{Div}\Xi = \text{Div} F$ with $L$ given by \eqref{lagrangian} the result would be negative. This situation is clearly depicted in the Appendix.

If we invert the definitions for $\bar{\pi}_\alpha$ and use \eqref{classol} in $\bar{Q}_I$ we calculate the integrals' values on the solution space:
\be \label{intval}
Q_1 =\bar{Q}_1 = 4\,c\,M \quad , \quad Q_2=\bar{Q}_2= \frac{4}{c} \quad , \quad Q_3=\bar{Q}_3 = 2 \, c
\ee
\par
As it is known the Schwarzschild solution involves only one essential constant, the mass $M$. The second constant appearing in the solution \eqref{classol}, \eqref{linel2} can be seen to be absorbable by a scaling of the time coordinate $t\rightarrow \frac{t}{c}$, allowing to set $c=1$ but not $c=0$.  So, on the solution space we can set $\bar{Q}_1=4M$, $\bar{Q}_2=4$, $\bar{Q}_3=2$. It is noteworthy that the values of the last two integrals of motion can be changed at will and that they do not involve the essential parameter $M$ which characterizes the geometry. The above argument relies on the underlying geometry.

But what if we were deprived of the line element and we were just given the dynamical system \eqref{lagham}? How could we differentiate between the constants $\kappa_1$ and $\kappa_2$, $\kappa_3$? Interestingly enough, there is an argument that leads to a distinction between them: The crucial observation is that $\bar{Q}_1$ has a vanishing Poisson bracket with each kinetic term of $\bar{H}$, while $\bar{Q}_2$, $\bar{Q}_3$ need the entire kinetic term in order to produce vanishing Poisson brackets. This fact is reflected in the following property concerning the one-parameter family of canonical transformations generated by the charge $Q_1$ (see \cite{Pons1}, \cite{struck} for the generalization of Noether symmetries for constrained systems and for Noether's theorem in phase space),
\begin{align}
a\rightarrow e^{\lambda} a,\, \bar{\pi}_a \rightarrow e^{-\lambda} \bar{\pi}_a,\, b\rightarrow e^{-\lambda} b,\, \bar{\pi}_b \rightarrow e^{\lambda} \bar{\pi}_b
\end{align}
Under such transformations $\bar{Q}_1$ remains, of course, unchanged, while $\bar{Q}_2$, $\bar{Q}_3$ are scaled by $e^{-\lambda}$ and $e^{\lambda}$ respectively. One can thus use the freedom of $\lambda$ to arbitrarily change the values $\kappa_2$, $\kappa_3$, but not $\kappa_1$. Furthermore, the Hamiltonian $\bar{\mathcal{H}}_c$ remains, due to the particular scaling of $\bar{Q}_2,\, \bar{Q}_3$, unchanged.

If we return to the phase space, we can write
\be
 \label{M}
  \bar{Q}_{23} = \bar{Q}_2 \, \bar{Q}_3 =-\frac{1}{2\,b^2}\bar{\pi}_\alpha^2+\frac{1}{a\,b}\,\bar{\pi}_\alpha\,\bar{\pi}_\beta
\ee
with the relevant Poisson bracket algebra now becoming
\begin{subequations}
\begin{align} \label{poisson1}
  & \{\bar{Q}_{23}, \bar{\mathcal{H}}_c\} = 0 \\ \label{poisson2}
  & \{\bar{Q}_I , \bar{Q}_{23} \} = 0 ,\quad\quad I=1,2,3.
\end{align}
\end{subequations}
As expected $\bar{Q}_{23}$ is a quadratic integral of motion. From a group theoretical view, $\bar{Q}_{23}$ is an element of the centre of the universal enveloping algebra (uea) generated by $\bar{Q}_I$'s, i.e. it is the Casimir invariant. The Hamiltonian $\bar{\mathcal{H}}_c$ belongs also to the centre of uea (it commutes with all $\bar{Q}_I$'s), thus it can only differ from $\bar{Q}_{23}$ by an additive and/or a multiplicative constant. It is an easy matter for one to check that indeed
\be \label{Q23con}
\bar{Q}_{23}=8 \, (\bar{\mathcal{H}}_c + 1).
\ee

To sum up, we have constructed a gauge independent, quadratic in the momenta, integral of motion which commutes with the only integral of motion that entangles the sole essential constant of the Schwarzschild solution. In the next section, and in order to proceed with the quantization, we will rely on these two quantities.

\section{Quantization}
In order to quantize our system, we must turn into operators $\bar{\mathcal{H}}_c=\frac{1}{8}\left(\bar{Q}_{23}-8\right)$, $\bar{Q}_1$, $\bar{Q}_2$ and $\bar{Q}_3$ (hereafter, for the shake of simplicity, we will omit the $bars$ from the symbols of the corresponding operators). The corresponding quantum operators can be inferred from \eqref{quantcon}, \eqref{qlin}, \eqref{gbar} and \eqref{xiI}. On account of our previous general discussion, the measure entering the quantum operators ought to be taken as $\mu(a,b)= \lambda \sqrt{det|\bar{G}_{\alpha\beta}|}\propto a\,b$. The constancy of $\lambda$ is forced by the combined requirement that the $Q_I$'s must be realized as Hermitian operators and at the same time retain their classical geometrical character by acting as derivatives. Thus the extra term $\xi^\alpha_I \partial_\alpha \lambda$ must vanish for all $I=1,2,3$, which leads to a constant $\lambda$. Two further arguments in favor of this choice of measure are:
\begin{itemize}
\item The fact that the quantum analogue of the algebra \eqref{poisson2} is made isomorphic to the classical, i.e.
\be \label{quantalg}
[\widehat{Q}_I, \widehat{Q}_{23}]F(a,b)=0, \quad I=1,2,3 \quad \text{for any } F(a,b)
\ee
a fact that is highly non trivial, since it depends on the choice of both the factor ordering and the measure.

\item  At the classical level, the only linear integral of motion involving the essential constant is $Q_1$. If we seek the functions on the configuration space which are invariant under the point transformations generated by $Q_1$ we find $\{Q_1,f(a,b)\}=0 \Rightarrow f(a,b)=f(a\,b)$.
\end{itemize}
The above arguments lead to the following linear operators corresponding to the elements of the classical algebra, the Casimir invariant and the Hamiltonian:
\begin{align}
\widehat{Q}_1 &= -\ima \left(b\, \partial_b - a\, \partial_a\right) \\
\widehat{Q}_2 &= -\frac{\ima}{a\, b} \partial_a \\
\widehat{Q}_3 &= -\ima \left(\partial_b - \frac{a}{2\, b}\,\partial_a\right) \\
\widehat{Q}_{23} &= \frac{2}{b^2}\, \partial_a\partial_a -\frac{1}{a\, b}\, \partial_a\partial_b+\frac{1}{2 a\, b^2}\, \partial_b \\
\widehat{\mathcal{H}}_c & =\frac{1}{8}\left(\widehat{Q}_{23} - 8\right)
\end{align}
It is an easy task to check that these operators satisfy not only the relations
\begin{equation}
[\widehat{Q}_I, \widehat{Q}_J]F(a,b) =\ima\, C^{K}_{\phantom{K}IJ} \widehat{Q}_K F(a,b)
\end{equation}
for any test function $F(a,b)$, but also \eqref{quantalg} as well.

Due to the constraint condition \eqref{1st int}, applied to the specific structure constants inferred from \eqref{poisson_int_motion2}, we conclude that only the eigenvalue $\kappa_1$ is free and $\kappa_2$, $\kappa_3$ must necessarily be zero. However, the latter is impossible since on the classical solution space \eqref{intval} hold. The reasoning given below \eqref{1st int} results in the need to consider the two- and/or one- dimensional subalgebras. The investigation of these cases can be easily carried out.

As far as the $2d$ subalgebras are concerned the results obtained are briefly the following:
\begin{enumerate}[(a)]
 \item For the two non Abelian subgroups either $\kappa_2$ or $\kappa_3$ are forced to be zero, something that is inconsistent with their classical values.

\item For the Abelian subgroup, the two linear equations lead to the solutions $\Psi(a,b)= A\, \exp\left(\frac{\ima}{2} (\kappa_2\, a^2\, b+2\kappa_3\, b)\right)$, where $A$ is constant and the quadratic constraint enforces the restriction $\kappa_2\, \kappa_3=8$. It is of course doubtful if one can accept such a wave function to represent the geometry, knowing that it does not contain the essential constant $M$. However, one could interpret it as plane waves representing the limiting flat space-time $M=0$.
\end{enumerate}
The one-dimensional subalgebras spanned by $\widehat{Q}_2$, $\widehat{Q}_3$ give solutions which are special cases of the solution described in b), as expected since they commute.

Consequently, the only possibility is to adopt $\widehat{\mathcal{H}}_c$ and $\widehat{Q}_1$ as conditions on the wave function. This is indeed possible, since they commute with each other and therefore can be considered as physical quantities on the phase space that can be measured ``simultaneously" (our dynamical parameter is the distance $r$). The ensuing eigenvalue equations are

\begin{subequations} \label{Qu:1}
\begin{align}
\widehat{\bar{\mathcal{H}}}_c \Psi & = 0\Rightarrow a\,\partial_a\partial_a\,\Psi-2\,b\,\partial_a\partial_b\,\Psi+\partial_a\,\Psi-16\,a\,b^2\,\Psi=0 \label{Qu:1a} \\
\widehat{Q}_1 \Psi & = \kappa_1 \, \Psi\Rightarrow \ima\left(-a\,\partial_a\,\Psi+b\,\partial_b\,\Psi\right)=\kappa_1\, \Psi\label{Qu:1b}
\end{align}
\end{subequations}
The solution of the linear partial differential equation \eqref{Qu:1b} is
\begin{align}
\Psi(a,b)=a^{\ima\,\kappa_1}\,S(a\,b).
\end{align}
If we insert the above solution into the Hamiltonian constraint \eqref{Qu:1a} we arrive at the following ordinary differential equation for $S(u)$ $(u=a\,b)$:
\begin{align}\label{eqS}
u^2\,S''(u)+u\,S'(u)+\left(\kappa_1^2+16\,u^2\right)\,S(u)=0
\end{align}
which has the general solution
\begin{align}\label{final S}
S(u)=c_1\,J_{\ima\,\kappa_1}(4\,u)+c_2\,Y_{\ima\,\kappa_1}(4\,u),
\end{align}
in terms of the Bessel functions of imaginary order.

In order to gain some insight on the normalizability of the formal probability, instead of these Bessel functions and because of their imaginary order, we can use the functions $F_{\ima\, \kappa_1}(4\,u)$ and $G_{\ima\, \kappa_1}(4\,u)$ defined in \cite{Dunster} through the Hankel functions $H_\mu^{(1)}(u)=J_\mu(u)+\ima\, Y_\mu(u)$ and $H_\mu^{(2)}(u)=J_\mu(u)-\ima\, Y_\mu(u)$, $\mu \in \mathbb{C}$. Thus, the solution can be written as
\be
S(u)=c_1\,F_{\ima\, \kappa_1}(4\,u)+c_2\,G_{\ima\, \kappa_1}(4\,u)
\ee
with
\begin{align}
&F_{\ima\, \kappa_1}(4\,u) = \frac{1}{2} \left(e^{-\kappa_1\pi/2}\, H_{\ima\kappa_1}^{(1)}(4u)+ e^{\kappa_1\pi/2}\, H_{\ima\kappa_1}^{(2)}(4u)\right)\\
&G_{\ima\, \kappa_1}(4\,u) = \frac{1}{2\ima} \left(e^{-\kappa_1\pi/2}\, H_{\ima\kappa_1}^{(1)}(4u)- e^{\kappa_1\pi/2}\, H_{\ima\kappa_1}^{(2)}(4u)\right).
\end{align}
These functions are linearly independent solutions of \eqref{eqS} and have the following properties: a) when $u \in (0,+\infty)$ they are real, b) they are oscillatory with a phase difference of $\frac{\pi}{2}$ and c) when both $u$, and/or $\kappa_1$ tend to zero, $F_{\ima\, \kappa_1}(4\,u)$ tends to $1$, while $G_{\ima\, \kappa_1}(4\,u)$ becomes infinite.

The final form of the wave function $\Psi(a,b)$ is
\begin{align}\label{final psi}
\Psi(a,b)= a^{\ima\,\kappa_1} \, S(a\, b),
\end{align}
so we can define a probability density of the form
\be
\mu(a\, b)\, \Psi^*(a, b) \, \Psi (a, b) \propto u \, S^*(u)\,S(u).
\ee

\section{Discussion}
In this paper we give a definition of conditional symmetries \eqref{defcon} in terms of simultaneous conformal Killing fields $\xi_I$ of the minisuperspace metric $G^{\alpha\beta}$ and the potential $V$. We prove that these symmetries remain form invariant under a rescaling of the lapse function $N=\omega(q)\bar{N}$. Accordingly, we observe that there is a special scaling of the lapse by the inverse of the potential $\omega=\frac{1}{V}$, in which the form invariant $\xi_I$'s become Killing vector fields of $\bar{G}^{\alpha\beta}$. In this conformal gauge we show that these $\xi_I$'s are indeed equivalent to the variational (Noether) symmetries one would have found if one had used the standard definition of variational symmetries on the specially scaled Lagrangian $\bar{L}=\frac{1}{2\bar{N}}\bar{G}_{\alpha\beta}\dot{q}^\alpha\dot{q}^\beta-\bar{N}$. We also show that it is possible to find the same variational (Noether) symmetries in the arbitrary parametrization $L=\frac{1}{2N}\,G_{\alpha\beta}\dot{q}^\alpha\dot{q}^\beta-N\, V$ if \eqref{Lagcon} is used in the configuration space.

As a result, we propose a method to quantize the \emph{minisuperspace} actions which are described by the singular Lagrangian \eqref{Lag}. The steps of this procedure are:
\begin{enumerate}
\item Go over to the Hamiltonian $\bar{H}$.
\item Calculate the Noether symmetries as Killing fields of the metric $\bar{G}^{\alpha\beta}(q)$ and, trivially, symmetries of the constant potential.
\item Identify the essential constants of the metric via the arguments following \eqref{1st int}. Promote the allowed $Q_I$'s to operators according to \eqref{qdef} with $\mu=|\bar{G}|^{1/2}$.
\item Promote the Hamiltonian constrain $\bar{\mathcal{H}}_c$ (which is a linear function of the Casimir invariant) via  \eqref{quantcon} to a Hermitian operator acting on the wave function.
\end{enumerate}

In sections $3$ and $4$ we present an application of the above method for the case of static, spherically symmetric geometries. First, we begin from the Lagrangian \eqref{lagrangian} emanating from the line element \eqref{metric}. We find the simultaneous conformal Killing fields \eqref{xiI} of the supermetric and the potential, which define the three conserved charges \eqref{int_motion}. The unique Casimir invariant of their algebra is $Q_{23}$. In order to make it numerically proportional to the kinetic part of the Hamiltonian we are led to \eqref{lagham}.

\begin{itemize}
\item In \cite{Lamb} Capozziello and Lambiase use the standard Noether symmetries approach for the \emph{regular} system obtained by gauge fixing the lapse function occurring in a singular minisuperspace Lagrangian. Furthermore, they propose the use as quantum operators of as many of the symmetries as they can be simultaneously brought into normal form by a single coordinate transformation of the configuration space variables, so that they become manifestly \emph{cyclic}. They thus, effectively, invoke the maximal Abelian subgroup as the relevant tool for quantization. Our perspective is quite different, since we start with a singular Lagrangian \eqref{Lag} and we use only those symmetries that are allowed by the condition \eqref{1st int}. In this sense, we have refined their method and their search for "... a criterion by which the
Hartle point of view can be recovered without arbitrariness"; the selection rule \eqref{1st int} is exactly of that nature. An other worth-emphasizing point is that the use of the Noether symmetries must be restricted by the requirement that they correspond to \emph{essential} constants of the underlying geometry; in our example, the use of the maximal Abelian subgroup leads to the \emph{marginally} acceptable plane-wave solutions not containing the classical geometry's \emph{essential} constant $M$.
\item Our specific example (described in sections 3 and 4) has also been the subject of \cite{Vakili1}. In that work, Vakili finds two of the $Q_I$'s but then he uses a linear combination of them in order to reproduce the essential constant $M$ of the Schwarzchild metric. He thus reaches to the unique acceptable linear quantum operator equivalent to our $\widehat{Q}_1$. The clever choice of the lapse function, his equation (4), along with the somewhat unorthodox choice of factor ordering for the operators (see below his equation (55)) leads essentially to a constant potential (his equation (53)) and to the Laplacian operator (his equation (56)). As a result, the solution spaces found both by us and Vakili essentially coincide. Of  course, our general theory constitutes a systematic explanation of the various choices of his work.
\item In \cite{Pons} Jizba and Pons use Noether symmetries in order to transform a regular Lagrangian into a singular one: they promote, at the classical level, the constants of motion to constraints by adding them, with appropriate Lagrange multipliers, to the Lagrangian. The most natural step after this, in order to quantize the theory \emph{\'{a}la} Dirac, is to promote the constants of motion to operators acting on the wave function. This is exactly our way of thinking since we already start with a singular Lagrangian and we apply the appropriate $\widehat{Q}_I$ on the wave function.
\end{itemize}

A further point that we would like to stress is the somewhat unexpected result that the quantum algebra \eqref{1st int} of the linear operators, turns to be isomorphic to the classical one \eqref{classalg} by virtue of the form of the operators \eqref{qdef} and the relation among the $\xi$'s \eqref{comxi} inferred from \eqref{classalg}.

We plan to return with an exhaustive list of applications to all $2d$ and/or $3d$ configuration minisuperspaces, emanating from the appropriate spacetime geometries, i.e. Bianchi types.

\newpage
\appendix
\section{Appendix: Calculation of variational symmetries}

As already stated in the main text, if one searched for the generators $X_I$ of variational symmetries of \eqref{lagrangian} by using the relation
\be \label{varsymcon}
pr^{(1)}X_I(L) = 0,
\ee
none of the three $\xi$'s would emerge. On the other hand, condition \eqref{varsymcon} works for the reparametrised Lagrangian \eqref{lagham} and reveals the three conditional symmetries. This is due to the fact that we are dealing with a singular Lagrangian. In the first case, the $Q_I=\xi_I^\alpha \pi_\alpha$ are integrals of motion on the reduced phase space $\mathcal{H}_c\approx 0$, while in the latter, $\{Q_I,H\}=0$.

In order to calculate the right $X_I$'s in any reparametrisation we use the definition \eqref{Lagcon} in which there exists an extra term proportional to the quadratic constraint equation involving the velocities
\be \label{varsymextra}
pr^{(1)}X_I(L) = \phi (a,b)\,n\, EL(n)
\ee
In the case of the Schwarzschild example we have
\begin{align}
X & := \eta_a (a,b) \frac{\partial}{\partial a}+\eta_b (a,b) \frac{\partial}{\partial b} \\
pr^{(1)} X & := X + \frac{d\eta_a}{dr} \frac{\partial}{\partial a'} + \frac{d\eta_b}{dr} \frac{\partial}{\partial a'}
\end{align}
and by substituting \eqref{lagrangian} inside \eqref{varsymextra} it is easy to calculate that
\be
X_1 = -a \frac{\partial}{\partial a} + b \frac{\partial}{\partial b}, \quad
X_2 = \frac{1}{a b} \frac{\partial}{\partial a}, \quad
X_3 = -\frac{a}{2b} \frac{\partial}{\partial a} +  \frac{\partial}{\partial b},
\ee
which are exactly the $\xi_I$'s of \eqref{xiI}. The corresponding functions $\phi_I$ that multiply the Euler - Lagrange equation $EL(n)=\frac{\partial L}{\partial n}$ are
\be
\phi_1 =  -1, \quad \phi_2 =\frac{1}{a^2 b}, \quad \phi_3 = - \frac{1}{2 b}
\ee
and are equal to the multiplying factors in \eqref{poisson_int_motion} divided by $-n$.

As a concluding remark it is reasonable to say that, it is of utmost importance not to fix the gauge, i.e. set $n=1$, since the presence of the quadratic constraint is needed in order to acquire all the variational symmetries regarding a singular Lagrangian. This, of course, does not mean that one is prohibited to select a gauge for $n$; one must simply remember to take into account the gauge fixed form of the constraint equation as in \eqref{varsymextra}. The extra term in \eqref{varsymextra} can be interpreted as a component of a more general generator, $\widetilde{X}_I=-\phi_I\,n\, \partial_ n+ X_I$, expressing not only transformations of the reduced configuration space $(a,b)$ as $X_I$'s do, but transformations over the full space $(n,a,b)$. This does not add any term of the form $\frac{\partial}{\partial n'}$ to the prolonged vector, since the Lagrangian is free of $n'$.

\end{document}